\documentclass[bibnotes,amsmath,amssymb,showpacs,floatfix,superscriptaddress,reprint]{revtex4-1}
\usepackage{amsmath,empheq}
\usepackage{amsfonts}
\usepackage{amssymb}
\usepackage{amsxtra}
\usepackage{mathtools}
\usepackage{xcolor}
\usepackage{graphicx}
\usepackage{subfigure}
\usepackage{dcolumn}
\usepackage{mathrsfs}
\usepackage{float}
\usepackage{bm}
\usepackage[breaklinks=true,colorlinks,citecolor=blue,linkcolor=blue,urlcolor=blue]{hyperref}

\DeclareMathAlphabet{\bi}{OML}{cmm}{b}{it}
\def\be{\begin{equation}}
\def\ee{\end{equation}}
\def\bearr{\begin{eqnarray}}
\def\eearr{\end{eqnarray}}

\def\bs{\boldsymbol}

\begin{document}
\title{Hyperbolic polaritons in topological nodal ring semimetals}
\author{Ashutosh Singh}
\email{asingh.n19@gmail.com}
\affiliation{Department of Physics and Astronomy, Texas A\&M University, College Station, TX, 77843 USA}
\affiliation{Department of Physics, National Tsing Hua University, Hsinchu 300013, Taiwan}
\author{Maria Sebastian}
\affiliation{Department of Physics and Astronomy, Texas A\&M University, College Station, TX, 77843 USA}
\author{Yuanping Chen}
\affiliation{School of Physics and Electronic Engineering, Jiangsu University, Zhenjiang, 212013, Jiangsu, China}
\author{Po-Yao Chang}
\email{prayser@gmail.com}
\affiliation{Department of Physics, National Tsing Hua University, Hsinchu 300013, Taiwan}
\author{Alexey Belyanin}
\email{belyanin@tamu.edu}
\affiliation{Department of Physics and Astronomy, Texas A\&M University, College Station, TX, 77843 USA}

\date{\today}

\begin{abstract}
In mirror-symmetric systems, there is a possibility of the realization of extended gapless electronic states characterized as nodal lines or rings. Strain induced modifications to these states lead to emergence of different classes of nodal rings with qualitatively different physical properties. Here we study optical response and the electromagnetic wave propagation in  type I nodal ring semimetals, in which the low-energy quasiparticle dispersion is parabolic in momentum $k_x$ and $k_y$ and is linear in $k_z$. This leads to a highly anisotropic dielectric permittivity tensor in which the optical response is plasmonic in one spatial direction and dielectric in the other two directions. The resulting normal modes (polaritons) in the bulk material become hyperbolic over a broad frequency range, which is furthermore tunable by the doping level. The propagation, reflection, and polarization properties of the hyperbolic polaritons not only provide valuable information about the electronic structure of these fascinating materials in the most interesting region near the nodal rings but also pave the way to tunable hyperbolic materials with applications ranging from anomalous refraction and waveguiding to perfect absorption in ultrathin subwavelength films.
\end{abstract}

%
\maketitle
{\it\underline{Introduction}:}
The quantification of topological properties of condensed matter systems in the last decade has been driven to a large extent by the studies of Dirac and Weyl semimetals \cite{Ashvin,burkov1}. In these materials, the conduction and the valence bands merge at isolated points in the Brilliouin zone such that the low-energy quasiparticles mimic the physics of Dirac and Weyl fermions with speed much lower than light. Low-energy optical spectroscopy provides a unique opportunity for their energy-resolved studies  near band crossings, which is not always possible by other means. Perhaps the most direct consequence of the Weyl fermion dispersion is a linear in frequency conductivity \cite{PhysRevB.92.075107, PhysRevB.92.241108,PhysRevB.93.121110,PhysRevB.93.121202,PhysRevB.87.235121}, with modifications due to anisotropic dispersion \cite{Jules} band with temperature playing important role due to the quadratic dependence of density of states on quasiparticle energy \cite{singh1,singh2}. A lot of effort has been spent on extracting the topological features of these materials from their optical properties; see, e.g.,  \cite{long2018,rostami2018,belyanin,chen2019,moore2019} and references therein.

\begin{figure}[ht!]
\includegraphics[width = 0.9\linewidth]{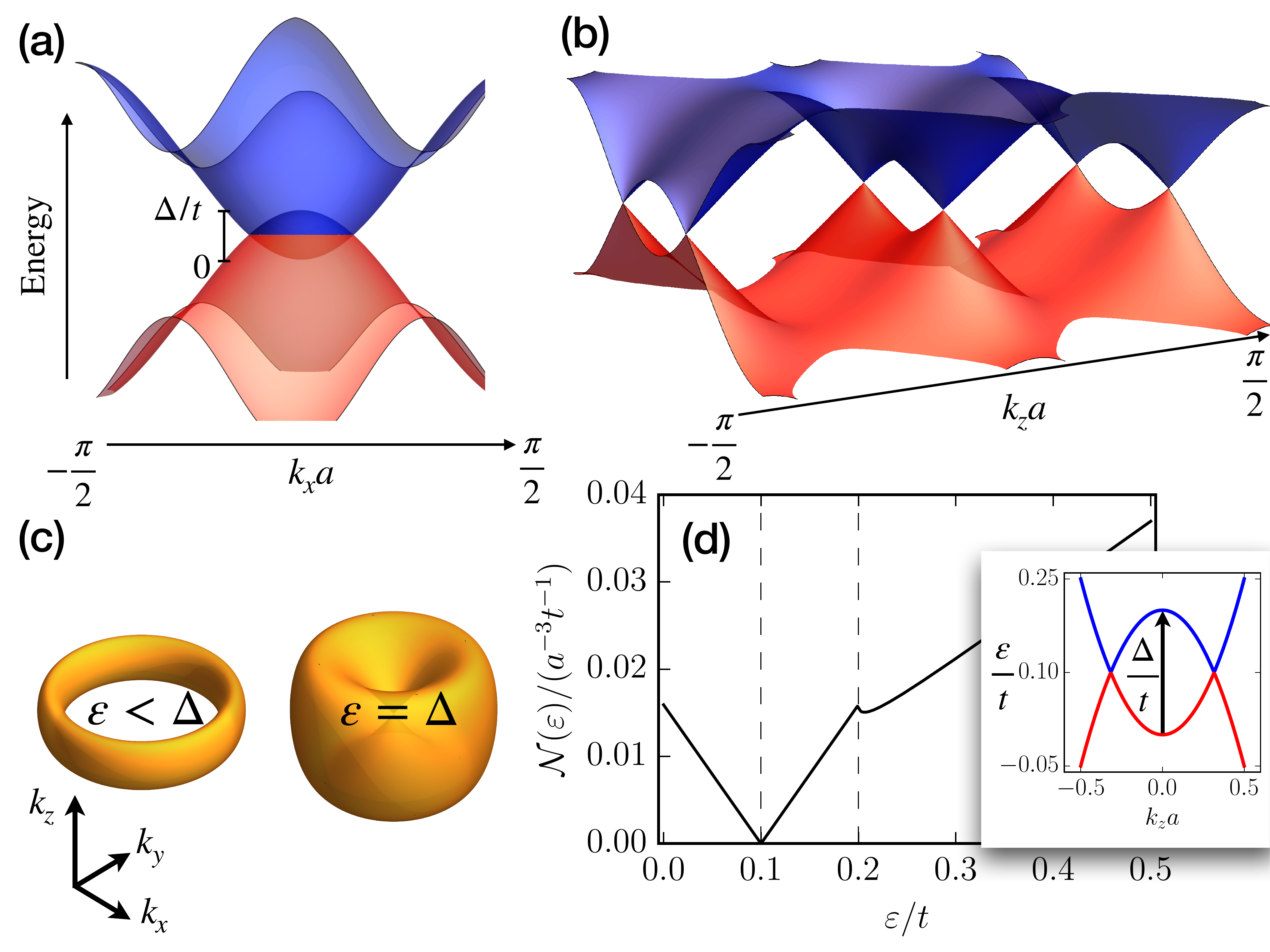}
\caption{Energy dispersion in (a) $k_z=0$, and (b) $k_y =0$ momentum planes for type-I NRSM described by the Hamiltonian in Eq.~\eqref{Hamiltonian_TB}. The vertical axis is normalized by $t$. (c) Constant energy surfaces. For energies lower than $\Delta$ the momentum distribution forms a toroidal shape. Increasing energy deforms the toroid and it collapses into a drum-like structure for energies greater than $\Delta$. (d) The density of states, ${\mathcal N}(\varepsilon)$, normalized by $a^{-3}t^{-1}$ as a function of energy $\varepsilon$ normalized by $t$ for $\Delta = 0.2 t$.}
\label{dispersion}
\end{figure}

In a rather new class of topological semimetals known as nodal line semimetals, the conduction and valence band touch along a line or a ring (loop) \cite{Fang_2016}. Different classes of nodal rings have been proposed, e.g., hybrid nodal rings \cite{PhysRevMaterials.6.034202}, spin gapless nodal rings \cite{YANG202143}, topological nodal rings in carbon networks \cite{PhysRevB.97.121108}, antiperovskites \cite{PhysRevLett.115.036807}, semimetallic carbon tetrarings \cite{Puru}, and orthorhombic $C_{16}$\cite{PhysRevLett.116.195501}. Among many interesting features displayed by the nodal ring semimetals (NRSM) are unusual Landau level quantization \cite{PhysRevB.92.045126}, and drumhead surface states \cite{PhysRevB.93.205132, PhysRevB.93.121113}. 
Furthermore, the bulk energy dispersion is highly anisotropic in momentum space as shown in Fig.~\ref{dispersion}(a) and Fig.~\ref{dispersion}(b). 
Moreover, 
abrupt change in Fermi surface topology occur when the quasiparticle energy is tuned in the vicinity of the energy gap parameter (Fig.~\ref{dispersion}(c)). Direct consequence of this feature appears in the density of states (DOS), Fig.~\ref{dispersion}(d). 

One can fully expect that these unusual electronic properties of NRSM result in a peculiar and even unique optical response. Previous studies were mainly focused on the derivation of the linear optical conductivity spectra \cite{carbotte2017,barati2017,pronin2021} as well as the second-order conductivity in symmetry-broken NRSM \cite{zuber2021}. However, the aspect of the optical response which provides most insight into the physical properties, and also the one most closely connected to experiment is the propagation, absorption, reflection/refraction, and polarization properties of the normal EM modes of the material, or the polaritons. In this paper we focus on type I NRSM in which the connection between the fascinating properties of the polaritons and the underlying electronic structure is very intuitive. One obvious property of the polaritons in NRSM stems from the fact that 
the low-energy quasiparticle dispersion is parabolic in momentum $k_x$ and $k_y$ but is linear in $k_z$. This leads to uniaxial anisotropy of the dielectric permittivity tensor in which the optical response is plasmonic in one spatial direction and dielectric in the other two directions. The resulting polaritons in the bulk material split into so-called ordinary and extraordinary modes, and the extraordinary mode become hyperbolic over a broad frequency range, which can easily extend to 1-2 eV and which is furthermore tunable by the doping level. Note that in ``conventional'' Weyl semimetals with nodal points the hyperbolic dispersion was only predicted in high magnetic fields and with Fermi level tuned to the band crossing points \cite{long2018}. Note also that in a relatively better studied group of open nodal line semimetals the hyperbolic dispersion has been recently observed with tip-enhanced infrared spectroscopy  \cite{basov2022}. In other kinds of anisotropic crystals,  such as hexagonal boron nitride, the hyperbolic dispersion typically exists in a narrow mid-infrared frequency range defined by separation between anisotropic phonon resonances \cite{basov2023}. The hyperbolic materials are of course highly desirable for applications as they exhibit a plethora of unique properties such as negative refraction, propagation through subwavelength apertures, and waveguiding by ultrathin films.  

The existence of two types of polaritons and their hyperbolic character defines all aspects of the EM wave interaction with NRSM. Here we only briefly describe a few of them, hoping to stimulate subsequent studies and experiments. 


%
{\it\underline{Electron states in NRSM}}:
An effective low-energy Hamiltonian which describes different types of the NRSM  can be 
written as \cite{PhysRevB.97.121108} 
\begin{eqnarray}\label{Hamiltonian_TB}
H({\bf k}) = \left(\begin{array}{cc}
 t_1 {\mathscr G}(k_x,k_y) & it_2 \sin(k_z a) \\
 - it_2 \sin(k_z a) & \Delta + \gamma t_1 {\mathscr G}(k_x,k_y) \\
\end{array}
\right)~,
\end{eqnarray}
where ${\mathscr G}(k_x,k_y) = 2-\cos(k_x a)-\cos(k_ya)$, $t_1$ and $t_2$ are the hopping parameters, $a$ is the lattice spacing, $\Delta$ is the gap at the $\Gamma$ point and $\gamma$ is the band tuning parameter which takes value $-1$ for type-I NRSM and $0 < \gamma < 1$ for type-II NRSM. A third class of topological NRSM  comprises of merging type-I and type-II materials for which $\gamma$ as well as ${\mathscr G}(k_x,k_y)$ changes.
The nodal lines are protected by the mirror symmetry $M_z$, $M^{-1}_z H({\bf k})M_z = H(\bar {\bf k}) $ with $\bar {\bf k} = (k_x, k_y, -k_z)$ and $M_z = \sigma_z$.
One can visualize the nodal lines as the Berry flux tubes in the momentum space. These Berry flux tubes are robust objects due to quantization of the flux to integer multiples of $\pi$.

In this letter, we focus on type-I nodal rings with $\gamma=-1$, and we further take $t_1 = t_2 = t$ for simplicity. All results can be easily generalized for $t_1 \neq t_2$ if needed for specific compounds. One should expect the lattice constant $a$ to be of the order of 0.1-0.3 nm, whereas the hopping energy $t$ is typically on the scale of several eV. To fix the numerical value of the product $at$ for the plots, we assume that the ``Fermi velocity'' $v_F$, i.e., the linear slope of the electron dispersion in Fig.~\ref{dispersion}(b), satisfies $\hbar a^{-1} v_F = t$, whereas its ratio to the speed of light is $v_F/c = 300$. This is true within a factor of 2 for most Dirac materials. The parameter $\Delta$ could vary in wide limits. The most optimal situation for optical studies of topological nodal ring states is when $\Delta$ is small as compared to $t$, so that the nodal rings and characteristic optical transitions at photon energies $\sim \Delta$ are near the center of the Brillouin zone and well separated from higher-energy transitions between any trivial remote bands. As we see below, this will also maximize the optical anisotropy. We will set $\Delta = 0.2 t$ for further discussion. The corresponding electron energy dispersion is shown in Fig.~\ref{dispersion}. 


%
%
%
%
%
%

The quasiparticle energy dispersion for the Hamiltonian in Eq.\eqref{Hamiltonian_TB} is given as  
\begin{eqnarray}
\varepsilon_{\lambda\bf k} = \frac{\Delta}{2} +\frac{\lambda}{2}\sqrt{(\Delta -2 {\mathscr G}(k_x,k_y) t)^2+4 t^2 \sin ^2(k_z a)}~,
\end{eqnarray}
where $\lambda = +1 (-1)$ for conduction (valence) band. In the electric dipole approximation the  interband optical transitions are vertical. 
The transition energy for a quasiparticle at momentum ${\bf k}$ is given by the difference between the conduction and the valence band energies,
\begin{eqnarray}
\hbar \omega_{\bf k} = \sqrt{(\Delta -2 {\mathscr G}(k_x,k_y) t)^2+4 t^2 \sin ^2(k_z a)} ~.
\end{eqnarray}
The normalized eigenvectors are 
\begin{align}
|\Psi_{\lambda\bf k}\rangle = 
                          \frac{1}{{\mathscr N}_{\lambda}}\begin{pmatrix}
                          i (-\Delta +2 t{\mathscr G}(k_x,k_y)+\lambda\hbar\omega_{\bf k}) \\
                          2t \sin (k_z a)
                          \end{pmatrix}~,
\end{align}
%
%
%
with ${\mathscr N}_\lambda = \sqrt{(-\Delta +2 t{\mathscr G}(k_x,k_y) +\lambda\hbar\omega_{\bf k})^2+4 t^2 \sin ^2(k_za)}$.\\

\begin{figure}[ht!]
\centering
\includegraphics[width = \linewidth]{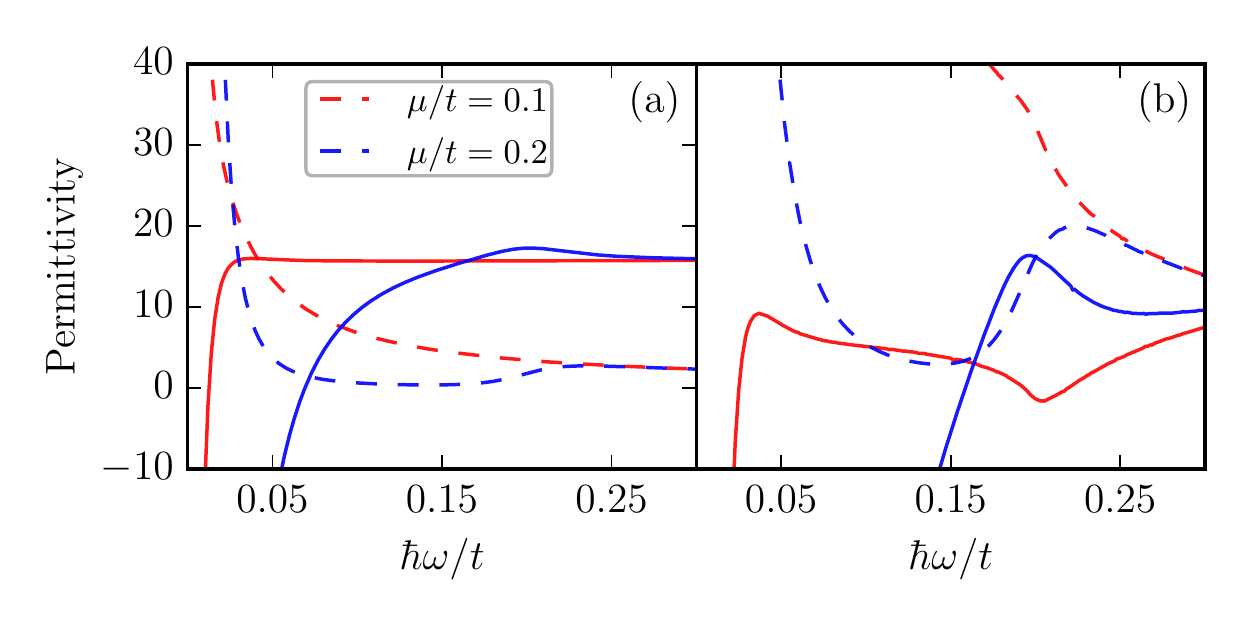}
\caption{ Real (solid line) and imaginary (dashed line) parts of (a) $\epsilon_{xx}$ and (b) $\epsilon_{zz}$ as a function of photon energy, for two different values of the chemical potential. The value of $\mu/t = 0.1$ corresponds to the chemical potential exactly at the band crossing, as one can see from Fig.~\ref{dispersion}(d). }
\label{fig2}
\end{figure}

%
%
{\it\underline{Optical permittivity}}~:
In equillibrium at temperature $T$ and chemical potential $\mu$, the linear response optical conductivity is computed within the Kubo framework~\cite{Mahan1990}. 
%
%
We will take $k_B T = t/200$ for numerical plots to include thermally excited carriers. 
In order to incorporate scattering-related losses at the phenomenological level, we have introduced a decay term, $\hbar\Gamma = 0.005t$. 
%
%
The current operator components  are $\hat j_{\alpha} = \hbar^{-1}\partial_{k_{\alpha}}\hat H$, where $\alpha = \lbrace x, y, z\rbrace$. For the Hamiltonian \eqref{Hamiltonian_TB} they become 
\begin{equation} 
\label{current} 
\hat {\bs j} = \frac{eat}{\hbar} \lbrace{\sin(k_xa)\hat\sigma_z, \sin(k_ya)\hat\sigma_z, -\cos(k_za)\hat\sigma_y\rbrace}.
\end{equation}
%
%
We also add the background permittivity ($\epsilon_b$) due to the sum of contributions from remote bands not included in the Hamiltonian \eqref{Hamiltonian_TB}, and assume it to be isotropic and with negligible dispersion within the frequency range of interest to us. Its exact value shifts the plots in Fig.~\ref{fig2} but does not change the qualitative physical behavior; we will use $\epsilon_b = 15$ as a reasonable number in the infrared. The resulting dielectric tensor $\hat\epsilon$ is expressed in terms of the conductivity (in SI units) as $\hat\epsilon(\omega) = \epsilon_b\mathbb{I}_{3\times3} + i\hat\sigma(\omega)/(\omega\epsilon_0)$.
%
%
Due to the symmetry of the system, only the diagonal terms of the conductivity tensor survive. The details of the conductivity derivation and analytic results 
are provided in the Supplementary Material (SM). 
The general structure and scaling of the diagonal permittivity components is given by $\epsilon_{\alpha \alpha}(\omega) = \epsilon_b - g\alpha_F c\mathscr{I}_{\alpha \alpha}/(2\pi^2a\omega)$,
%
%
where $\mathscr{I}_{\alpha \alpha}$ are dimensionless integrals specified in the SM, $\alpha_F = e^2\left(4\pi\epsilon_0\hbar c\right)^{-1}$ is the fine structure constant, and $c$ is the speed of light. 

%
%
%
%
%
To the leading order in the long wavelength limit, cylindrical symmetry is preserved so that  $\epsilon_{yy} \approx \epsilon_{xx}$. However, $\epsilon_{zz}$ behaves differently, as shown in Fig.~\ref{fig2}. First of all, the magnitude of the matrix elements of the $j_z$ component of the current is higher than the ones for $j_{x,y}$ components, as one can see from Eq.~\eqref{current} and the SM. Indeed, when $\Delta \ll t$ the main contribution comes from the states with $|k_{\alpha} a| \ll 1$ in the vicinity of the nodal rings. In this case the ratio of matrix elements $|j_z/j_{x,y}| \sim t/\Delta \gg 1$, yielding a higher magnitude of $\epsilon_{zz}$ as compared to $\epsilon_{xx}$. Second, while at the lowest frequencies all permittivity components are dominated by intraband plasmonic response (even when the Fermi level is at the band crossing energy, $\mu/t = 0.1$, because free carriers are still present at finite temperature), with increasing frequency the behavior of $\epsilon_{xx}$ becomes dielectric, whereas the $\epsilon_{zz}$ component maintains plasmonic behavior over a significantly broader frequency range. This extreme anisotropy with opposite signs of the real parts of the dielectric tensor components gives rise to the hyperbolic dispersion of the polaritons. 
 %
%
%
%
%
%
%

{\it\underline{Properties of NRSM polaritons}}:
\begin{figure}[]
\centering
\includegraphics[width = \linewidth]{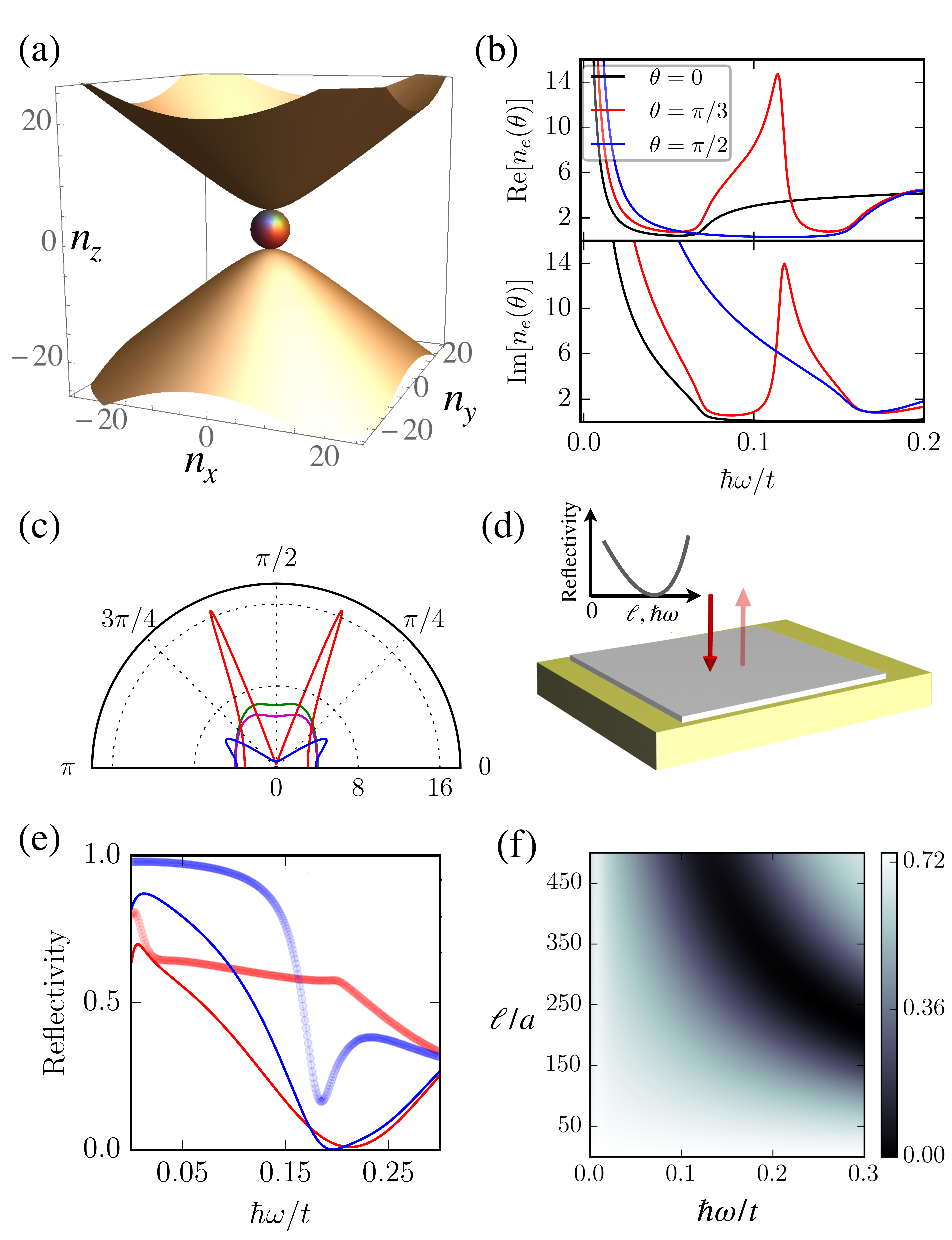}
\caption{ 
(a) The solution of the dispersion equation \eqref{disp1} for the  extraordinary wave (yellow hyperboloid) and  ordinary wave (red sphere) at a constant photon energy $\hbar\omega \sim 0.13 t$ and $\mu = 0.2 t$, so that $\epsilon_{xx} \sim 13$ and $\epsilon_{zz} \sim -20$. (b) ${\rm Re}[n_e]$ and ${\rm Im}[n_e]$ as a function of photon energy for $\mu = 0.2t$.
(c) Real part of $n_e(\theta)$ for $\mu = 0.1t, \hbar\omega = 0.1t$ (green), $\mu = 0.1t, \hbar\omega = 0.15t$ (purple), $\mu = 0.2t, \hbar\omega = 0.1t$ (red) and $\mu = 0.2t, \hbar\omega = 0.15t$ (blue).  (d) Schematic for the reflection of normally incident EM wave from an ultrathin NRSM film of thickness $\ell$ placed on top of a substrate of complex refractive index $n_d$. (e)  Ordinary (extraordinary) wave reflectivity shown in red line (circle) for $\mu = 0.1t$, and  in blue line (circle) for  $\mu = 0.2t$. The film thickness $\ell = 300a$. (f) Color plot of the ordinary wave reflectivity for $\mu = 0.1 t$ as a function of the thickness of the film (y-axis) and the photon frequency (x-axis). Here we assumed $n_d = 1.4+4.0i$. }
\label{fig3}
\end{figure}
Maxwell's equations for the electric field vector ${\bs E} \propto \exp(i {\bs q}{\bs r} - i \omega t)$ of monochromatic EM waves propagating in a bulk crystal with permittivity tensor $\hat\epsilon$ can be written as 
\bearr
\label{disp1} 
{\bs n}\left({\bs n}\cdot{\bs E}\right) - n^2{\bs E} + \hat\epsilon{\bs E} = 0, 
\eearr
where ${\bs n} = {\bs q}c/\omega$. For a diagonal permittivity tensor, the solution of the corresponding dispersion equation consists of two linearly polarized normal modes (polaritons) which are often called an ordinary and extraordinary wave. Since $\epsilon_{yy} = \epsilon_{xx}$, we can consider without loss of generality  the propagation with the wave vector in the $(xz)$-plane, i.e.,  ${\bs n} = ({n_x,0,n_z})$. Then the refractive indices of the ordinary and extraordinary modes are given by 
\begin{equation}
\label{n_sol}
n^2_o = \epsilon_{xx} ~\text{and}~
n^2_e = \frac{\epsilon_{xx}\epsilon_{zz}}{\epsilon_{xx}\sin^2\theta + \epsilon_{zz}\cos^2\theta}~,
\end{equation}
where $\theta = \cos^{-1}(n_z/|{\bs n}|)$. The electric field vector of the extraordinary mode lies in the $(xz)$-plane, whereas the one of the ordinary mode is along $y$. 

Figure~\ref{fig3}(a) shows an example of the constant-frequency surface for the dispersion equation of the two modes. The surfaces are plotted for $\mu = 0.2 t$ and the frequency $\hbar\omega \sim 0.13 t$ for which the real parts of the dielectric tensor components have a much greater magnitude than the imaginary parts, so that $\epsilon_{xx} \sim 13$ and $\epsilon_{zz} \sim -20$. For the ordinary waves the surface is a sphere which is a particular case of the usual Fresnel ellipsoid. At the same time, for the extraordinary modes the surface is a hyperboloid. Its cross-section at $n_y = 0$ is 
\begin{align}
\frac{n_x^2}{\epsilon_{zz}} + \frac{n_z^2}{\epsilon_{xx}} = 1. 
\end{align}
In the range of frequencies where Re$[\epsilon_{zz}] < 0$ and Re$[\epsilon_{xx}] > 0$ the EM waves are able to propagate in certain directions with $|{\bs n}| \gg 1$, i.e., $|{\bs q}| \gg \omega/c$, as one can also see in Fig.~\ref{fig3}(c). 

%
%
%

The dominant feature in the spectra of hyperbolic polaritons  is a characteristic peak in the extraordinary wave dispersion and absorption near the frequency which minimizes the denominator in the expression \eqref{n_sol} for $n_e^2$, see the spectra in Fig.~\ref{fig3}(b) for $\theta = \pi/3$. The resonance exists for any angle $\theta \neq 0$ or $\pi/2$. A similar phenomenon in classical anisotropic plasmas would be a hybrid plasmon-polariton resonance, corresponding to hybridization between longitudinal plasmons and transverse EM waves. Note also the existence of a photonic band gap at frequencies above the resonance, where the real part of the refractive index would have dropped to zero in the absence of an imaginary part of the permittivity tensor. The real part of the refractive index drops very steeply at the photonic band gap boundary, indicating a small group velocity $v_{group} \ll c$ in this region. This behavior is similar to the dispersion of extraordinary magnetopolaritons in nodal-point Weyl semimetals \cite{long2018}. 

Obviously, all of the above spectral and angular features in polariton propagation and absorption can have important practical applications in thin-film EM waveguides, modulators, switches etc. We will mention just one more potential application, which has been pointed out for strongly absorbing materials: ultrathin-film perfect absorbers \cite{rensberg2017}.  Consider an EM wave normally incident on an ultrathin (strongly subwavelength) NRSM film, as in Fig.~\ref{fig3}(d). In this case destructive interference between reflections from the front and back facets of the film can result in spectral windows of zero reflectivity even for a film much thinner than the incident wavelength. This is illustrated in Fig.~\ref{fig3}(e) for a fixed film thickness and in Fig.~\ref{fig3}(f) for a range of thicknesses of a few tens of nm, depending on the exact value of the lattice period $a$. The zero reflectivity region is tunable by doping, film thickness, and also depends on the substrate. As was shown in  \cite{rensberg2017}, the best results are obtained for metallic or highly doped semiconducting substrates with mostly imaginary refractive index, such as the one chosen for Fig.~\ref{fig3}(e,f).

In conclusion, topological nodal ring semimetals are natural hyperbolic optical materials, with associated extreme optical anisotropy, anomalous refraction, and strong plasmon-polariton resonances. Their unique combination of optical properties is highly sensitive to the material parameters and can be used for optical spectroscopy of the nodal rings. Ultrathin films of NRSM can find a number of applications as infrared waveguides, modulators, switches, and antireflection coatings. This work has been supported in part by the Air Force Office for Scientific Research Grant No.~FA9550-21-1-0272 and National Science Foundation Award No.~1936276. 
\bibliography{Ref2}
\end{document}